\documentclass[twocolumn, 9pt]{autart}

\usepackage{amssymb}
\usepackage{graphicx,bm}
\usepackage{enumitem}
\usepackage{dsfont, caption}
\usepackage[usenames, dvipsnames]{color}
\usepackage{todonotes}
\usepackage[colorlinks=true,
		raiselinks=true,
		linkcolor=MidnightBlue,
		citecolor=Mahogany,
		urlcolor=ForestGreen,
		pdfauthor={Sukumar Srikant and Debasish Chatterjee},
		pdftitle={A jammer's perspective of reachability and LQ optimal control},
		pdfkeywords={Pontryagin maximum principle, optimal control, adaptive control},
		pdfsubject={Manuscript},
		plainpages=false]{hyperref}
\usepackage{subfig}

\usepackage{mathtools}
\usepackage{txfonts}
\allowdisplaybreaks
\DeclareMathOperator*{\minimize}{minimize}
\DeclareMathOperator*{\argmax}{arg\,max}
\DeclareMathOperator*{\sbjto}{subject\ to}
\DeclareMathOperator{\Leb}{Leb}
\DeclareMathOperator{\sgn}{sgn}

\renewcommand{\mapsto}{\longmapsto}
\renewcommand{\le}{\leqslant}
\renewcommand{\ge}{\geqslant}

\newcommand{\R}{\mathds{R}}

\newcommand{\admact}{\mathds{U}}
\newcommand{\Let}{\coloneqq}
\newcommand{\teL}{\eqqcolon}
\newcommand{\dd}{\mathrm{d}}
\newcommand{\transp}{^\top}
\newcommand{\inverse}{^{-1}}
\newcommand{\opt}{^\ast}

\newcommand{\tinit}{\bar t}
\newcommand{\tfin}{\hat t}
\newcommand{\zinit}{\bar z}
\newcommand{\zfin}{\hat z}

\newcommand{\lra}{\longrightarrow}

\newcommand{\mcal}{\mathcal}

\newcommand{\placeqed}{\hfill{}\(\Box\)}

\newcommand{\norm}[1]{\left\lVert{#1}\right\rVert}
\newcommand{\abs}[1]{\left\lvert{#1}\right\rvert}
\newcommand{\Lp}[1]{\mathrm L_{#1}}
\newcommand{\indic}[1]{\mathds{1}_{#1}}
\newcommand{\inprod}[2]{\left\langle{#1}, {#2}\right\rangle}
\newcommand{\epower}[1]{\mathrm{e}^{#1}}
\newcommand{\pmat}[1]{\begin{pmatrix}#1\end{pmatrix}}
\newcommand{\secref}[1]{\S\ref{#1}}

\begin{document}

\begin{frontmatter}

\title{A jammer's perspective of reachability and LQ optimal control\thanksref{footnoteinfo}}
\thanks[footnoteinfo]{This paper was not presented at any IFAC meeting. Corresponding author D.\ Chatterjee. Tel. +91-22-2576-7879. Fax +91-22-2572-0057.}
\author[authors]{Sukumar Srikant}\ead{srikant.sukumar@iitb.ac.in},
\author[authors]{Debasish Chatterjee}\ead{dchatter@iitb.ac.in},
\address[authors]{Systems \& Control Engineering, Indian Institute of Technology Bombay, Powai, Mumbai 400076, India.}
          

\begin{abstract}
	This article treats two problems dealing with control of linear systems in the presence of a jammer that can sporadically turn off the control signal. The first problem treats the standard reachability problem, and the second treats the standard linear quadratic regulator problem under the above class of jamming signals. We provide necessary and sufficient conditions for optimality based on a nonsmooth Pontryagin maximum principle.
\end{abstract}

\end{frontmatter}

	\section{Introduction}
		Given a controllable linear system
		\[
			\dot x(t) = A x(t) + B u_1(t)
		\]
		with \(x(t)\in\R^d\) and \(u_1(t)\in\R^m\),\footnote{By controllability here we mean that the rank of the matrix \(\pmat{B & AB & \cdots & A^{d-1}B}\) is equal to \(d\).} we let a jammer corrupt the control \(u_1\) with a signal \(t\mapsto u_2(t)\in\{0, 1\}\) that enters multiplicatively, and that can sporadically be ``turned off'', i.e., set to \(0\). The effect, therefore, of \(u_2\) turning off is that the control \(u_1\) is deactivated simultaneously, and the system evolves in open-loop. The signal \(u_2\) provides a standard model for denial-of-service attacks for control systems in which the controller communicates with the plant over a network, and such models have been extensively studied in the context of cyberphysical systems; see, e.g., \cite{ref:RayMid-08} and the references therein. In this setting we ask whether it is possible to construct a control \(t\mapsto u_1(t)\) to execute the transfer of states of the resulting system 
		\begin{equation}
		\label{e:resulting system}
			\dot x(t) = A x(t) + B u_1(t) u_2(t)
		\end{equation}
		from given initial to given final states. Or, for instance, whether it is possible to stabilize the resulting system \eqref{e:resulting system} to the origin by suitably designing the control \(u_1\). Since both these problems are trivially impossible to solve if the jammer turns the signal \(u_2\) `off' entirely, to ensure a well-defined problem, in the adaptive control literature typically a persistence of excitation condition, such as, there exist \(T, \rho > 0\) such that for all \(t\) we have \(\frac{1}{T}\int_t^{t+T} u_2(s)\,\dd s \ge \rho\), is imposed on \(u_2\). Very little, however, is known about either reachability or stabilizability of \eqref{e:resulting system} under the above persistence of excitation condition. In particular, the problem of designing a state feedback \(u_1(t) \Let K(t) x(t)\) such that the closed-loop system is asymptotically stable under the preceding persistence of excitation condition, is open, with partial solutions reported in \cite{ref:SriAke-09}, \cite{ref:MazChiSig-13}.

		In this article we study two problems concerning the control system \eqref{e:resulting system}. In the first problem we turn the above-mentioned reachability question around and examine the limits of favourable conditions for the jammer. We ask the question: how long does the jamming signal \(u_2\) need to be set to `on' or \(1\) for the aforementioned reachability problem to be solvable? To wit, we are interested in the limiting condition such that if the jamming signal \(u_2\) is set to `off' or \(0\) for any longer time, then the standard reachability problem for \eqref{e:resulting system} under the control \(u_1\) would cease to be feasible. More precisely, we study the optimal control problem: given initial time \(\tinit\) and final time \(\tfin > \tinit\),
		\begin{equation}
		\label{e:opt reach problem}
		\begin{aligned}
			& \minimize_{u_1, u_2}	&& \norm{u_2}_{\Lp 0([\tinit, \tfin])}\\
			& \sbjto	\;	\quad	&& 
				\begin{cases}
					\dot z(t) = A z(t) + B u_1(t) u_2(t) \quad \text{for a.e.\ }t\in[\tinit, \tfin],\\
					z(\tinit) = \zinit\in\R^d, \quad z(\tfin) = \zfin\in\R^d,\\
					u_1:[\tinit, \tfin]\lra \admact\subset\R^m\text{ compact},\\
					u_2:[\tinit, \tfin]\lra\{0, 1\},\\
					u_1, u_2\text{ Lebesgue measurable},
				\end{cases}
		\end{aligned}
		\end{equation}
		Here the cost function is the \(\Lp 0\)-seminorm of the control \(u_2\), defined to be the Lebesgue measure of the set of times at which the control is non-zero, i.e.,
		\[
			\norm{u_2}_{\Lp 0([\tinit, \tfin])} \Let \Leb\Bigl( \bigl\{s\in[\tinit, \tfin]\,\big|\, u_2(s)\neq 0\bigr\}\Bigr).
		\]
		We assume that the time difference \(\tfin - \tinit\) is larger than the minimum time required to execute the transfer of the states from \(\zinit\) to \(\zfin\) in order to have a well-defined problem, and in addition assume that \(0\in\R^m\) is contained in the interior of \(\admact\). Notice that while the control \(u_1\) tries to execute the desired manoeuvre, the control \(u_2\) tries to switch to `on' for the least length of time to enable execution of the aforementioned manoeuvre. We provide necessary conditions for these reachability manoeuvres and in addition provide conditions for optimality in \eqref{e:opt reach problem}.

		The second problem that we study in this article is that of the performance of the linear quadratic regulator with respect to the control \(u_1\) in the presence of the jammer \(u_2\). We ask the question: How good is the performance of the standard linear quadratic regulator when the jammer corrupts the \(u_1\) signal by turning it `off' sporadically? To be precise, given symmetric and non-negative definite matrices \(Q_f, Q\in\R^{d\times d}\) and a symmetric and positive definite matrix \(R\in\R^{m\times m}\), initial time \(\tinit\) and final time \(\tfin > \tinit\), we study the following optimal control problem:
		\begin{equation}
		\label{e:opt lq problem}
		\begin{aligned}
			& \minimize_{u_1, u_2}
							&& \gamma \norm{u_2}_{\Lp 0([\tinit, \tfin])} + \tfrac{1}{2}\inprod{z(t)}{Q_f z(t)} \\
							& && \qquad + \tfrac{1}{2} \int_{\tinit}^{\tfin}\Bigl( \inprod{z(t)}{Q z(t)} + \inprod{u_1(t)}{R u_1(t)}\Bigr)\,\dd t\\
			& \sbjto\;	\quad	&&  
				\begin{cases}
					\dot z(t) = A z(t) + B u_1(t) u_2(t) \quad \text{for a.e.\ }t\in[\tinit, \tfin],\\
					z(\tinit) = \zinit\in\R^d,\\
					u_1:[\tinit, \tfin]\lra\R^m,\\
					u_2:[\tinit, \tfin]\lra\{0, 1\},\\
					u_1, u_2 \text{ Lebesgue measurable},
				\end{cases}
		\end{aligned}
		\end{equation}
		where \(\gamma > 0\) is a fixed constant. If \(u_2\) is set to `off' for the entire duration \([\tinit, \tfin]\), the cost accrued by the quadratic terms corresponding to an \(\Lp 2([\tinit, \tfin])\) cost involving the states \(z\) and the control \(u_1\) will be high. If \(u_2\) is set to `on' for the entire duration \([\tinit, \tfin]\), the cost corresponding to \(\norm{u_2}_{\Lp 0([\tinit, \tfin])}\) will be high. Any solution to the optimal control problem \eqref{e:opt lq problem} strikes a balance between the two costs: \(\Lp 2([\tinit, \tfin])\)-costs with respect to \(u_1\) and the states, and the \(\Lp 0([\tinit, \tfin])\)-cost with respect to \(u_2\). As in the case of \eqref{e:opt reach problem}, we provide necessary conditions for solutions to \eqref{e:opt lq problem}, and in addition provide sufficient conditions for optimality in \eqref{e:opt lq problem}.

		It turns out that the optimal control \(u_1\opt\) corresponding to the optimal control problem \eqref{e:opt reach problem} is the sparsest control that achieves the steering of the states from \(\zinit\) to \(\zfin\) within the allotted time \(\tfin - \tinit\) --- see Remark \ref{r:connection to maximally sparse}. The optimal control problem \eqref{e:opt lq problem} is closely related to the ``sparse quadratic regulator'' problem treated in \cite{ref:JovLin-13}; see Remark \ref{r:sparse quadratic regulator}. While the authors of \cite{ref:JovLin-13} approached the optimal control problem using approximate methods via \(\Lp 1\) and total variation relaxations, it is possible to tackle the problem directly without any approximations, as we demonstrate in Remark \ref{r:sparse quadratic regulator}. Sparse controls are increasingly becoming popular in the control community with pioneering contributions from \cite{ref:LinFarJov-11}, \cite{ref:JovLin-13}, \cite{ref:LinFarJov-13}, \cite{ref:FarLinJov-14}, \cite{ref:Bah-15}, \cite{ref:NagQueNes-16}, \cite{ref:IkeNag-14}, \cite{ref:NagQueNes-14}, \cite{ref:PolKhlSch-13}, \cite{ref:PolKhlSch-14}. Two distinct threads have emerged in this context: one, dealing with the design of sparse control gains, as in \cite{ref:Bah-15}, \cite{ref:PolKhlSch-13}, \cite{ref:PolKhlSch-14}, and two, dealing with the design of sparsest control maps as functions of time, as evidenced in the articles \cite{ref:JovLin-13}, \cite{ref:NagQueNes-14}, \cite{ref:IkeNag-14}, \cite{ref:NagQueNes-16}. With respect to \cite{ref:Bah-15}, \cite{ref:PolKhlSch-13}, \cite{ref:PolKhlSch-14} our work differs in the sense that we do not design sparse feedback gains, but are interested in the design of sparse control maps that attain certain control objectives. The articles \cite{ref:JovLin-13}, \cite{ref:NagQueNes-14}, \cite{ref:IkeNag-14}, \cite{ref:NagQueNes-16} deal with \(\Lp 0\)-optimal control problems, but none of them treat the precise conditions for \(\Lp 0\)-optimality, preferring instead to approximate sparse solutions with the aid of \(\Lp 1\)-regularized optimal control problems. To the best of our knowledge, this is the first time that the two optimal control problems \eqref{e:opt reach problem} and \eqref{e:opt lq problem} are being studied.

		Observe that both the optimal control problems \eqref{e:opt reach problem} and \eqref{e:opt lq problem} involve discontinuous instantaneous cost functions, and are consequently difficult to solve. We employ a nonsmooth version of the Pontryagin maximum principle to solve these two problems and study the nature of their solutions. Insofar as the existence of optimal controls is concerned, once again, the discontinuous nature of the instantaneous cost functions lends a nonstandard flavour to the above two problems. We derive our sufficient conditions for optimality with the aid of what is known as an inductive technique. These results are presented in \secref{s:main results}. We provide detailed numerical experiments in \secref{s:examples} and conclude in \secref{s:conclusion}.

		Our notations are standard; in particular, for a set \(S\) we let \(\indic{S}(\cdot)\) denote the standard indicator/characteristic function defined by \(\indic{S}(z) = 1\) if \(z\in S\) and \(0\) otherwise, and we denote by \(\inprod{v}{w} = v\transp w\) the standard inner product on Euclidean spaces.

    \section{Main Results}
	\label{s:main results}
		We apply the nonsmooth maximum principle \cite[Theorem 22.26]{ref:Cla-13} to the optimal control problems \eqref{e:opt reach problem} and \eqref{e:opt lq problem}, for which we first adapt the aforementioned maximum principle from \cite{ref:Cla-13} to our setting, and refer the reader to \cite{ref:Cla-13} for related notations, definitions, and generalizations:
		\begin{thm}
		\label{t:Clarke extended PMP}
			Let \(-\infty < \tinit < \tfin < +\infty\), and let \(\admact\subset\R^m\) denote a Borel measurable set. Let a lower semicontinuous instantaneous cost function \(\R^d\times\admact\ni(\xi, \mu)\mapsto\Lambda(\xi, \mu)\in\R\), with \(\Lambda\) continuously differentiable in \(\xi\) for every fixed \(\mu\),\footnote{Recall that a map \(\varphi:X\lra\R\) from a topological space \(X\) into the real numbers is said to be lower semicontinuous if for every \(c\in\R\) the set \(\{z\in X\mid \varphi(z) \le c\}\) is closed.} and a continuously differentiable terminal cost function \(\ell:\R^d\times\R^d\lra\R\) be given. Consider the optimal control problem
			\begin{equation}
			\label{e:Clarke extended opt problem}
			\begin{aligned}
				& \minimize_{u}	&& \quad  \ell\bigl(x(\tinit), x(\tfin)\bigr) + \int_{\tinit}^{\tfin} \Lambda\bigl(x(t), u(t)\bigr) \, \dd t \\
				& \sbjto		&& \quad  \begin{cases}
						\dot x(t) = f\bigl( x(t), u(t)\bigr)\quad \text{for a.e.\ }t\in[\tinit, \tfin],\\
						u(t) \in\admact\quad\text{for a.e.\ }t \in [\tinit, \tfin],\\
						u\text{ Lebesgue measurable},\\
						\bigl(x(\tinit), x(\tfin)\bigr) \in E \subset\R^d\times\R^d,
					\end{cases}
			\end{aligned}
			\end{equation}
			where \(f:\R^d\times\R^m\lra\R^d\) is continuously differentiable, and \(E\) is a closed set. For a real number \(\eta\), we define the \emph{Hamiltonian} $H^\eta$ by 
			\[
				H^\eta(x, u, p) = \inprod{p}{f(x, u)} - \eta \Lambda(x, u).
			\]
			If \([\tinit, \tfin]\ni t\mapsto \bigl( x\opt(t), u\opt(t) \bigr) \) is a local minimizer of \eqref{e:Clarke extended opt problem}, then there exist an absolutely continuous map \( p: [\tinit, \tfin] \lra \R^d \) together with a scalar \( \eta \) equal to \(0\) or \(1\) satisfying the \emph{nontriviality condition}
			\begin{equation}
			\label{e:Clarke:nontriviality}
				\bigl(\eta, p(t)\bigr) \neq 0 \quad \text{for all } t \in [\tinit, \tfin],
			\end{equation}
			the \emph{transversality condition}
			\begin{equation}
			\label{e:Clarke:transversality}
				\bigl(p(\tinit), - p(\tfin)\bigr) \in \eta \partial_x\ell\bigl(x\opt(\tinit), x\opt(\tfin)\bigr) + N_E^L \bigl(x\opt(\tinit), x\opt(\tfin)\bigr),
			\end{equation}
			where \(\partial_x\ell\) is the gradient of \(\ell\) and \(N_E^L\bigl(x\opt(\tinit), x\opt(\tfin)\bigr)\) is the limiting normal cone of \(E\) at the point \(\bigl(x\opt(\tinit), x\opt(\tfin)\bigr)\),\footnote{The \emph{limiting normal cone} of a closed subset \(S\) of \(\R^\nu\) is defined by means of a closure operation applied to the \emph{proximal normal cone} of the set \(S\); see, e.g., \cite[p.\ 240]{ref:Cla-13} for the definition of the proximal normal cone, and \cite[p.\ 244]{ref:Cla-13} for the definition of the limiting normal cone.} the \emph{adjoint equation}
			\begin{equation}
			\label{e:Clarke:adjoint}
				- \dot p(t) = \partial_x H^\eta \bigl( \boldsymbol\cdot, u\opt(t), p(t) \bigr) (x\opt(t))\quad\text{for a.e.\ }t\in[\tinit, \tfin],
			\end{equation}
			the \emph{Hamiltonian maximum condition}
			\begin{equation}
			\label{e:Clarke:Hamiltonian max}
				H^\eta\bigl( x\opt(t), u\opt(t), p(t)\bigr) = \sup_{v\in\admact} H^\eta\bigl( x\opt(t), v, p(t)\bigr)\quad\text{for a.e.\ }t\in[\tinit, \tfin],
			\end{equation} 
			as well as  the \emph{constancy of the Hamiltonian}
			\begin{equation}
			\label{e:Clarke:Hamiltonian constancy}
				H^\eta\bigl(x\opt(t), u\opt(t), p(t)\bigr) = h\quad\text{for a.e.\ }t\in[\tinit, \tfin] \text{ and some }h\in\R.
			\end{equation}
		\end{thm}

		The assumptions of \cite[Theorem 22.26]{ref:Cla-13} are considerably weaker than what we have stipulated above; we refer the reader to \cite[Chapter 22]{ref:Cla-13} for details.

		The quadruple \(\bigl(\eta, p(\cdot), x\opt(\cdot), u\opt(\cdot)\bigr)\) is known as the \emph{extremal lift} of the optimal state-action trajectory \([\tinit, \tfin]\ni t\mapsto \bigl(x\opt(t), u\opt(t)\bigr)\). The number \(\eta\) is called the \emph{abnormal multiplier}. The \emph{abnormal} case --- when \(\eta = 0\) --- may arise, e.g., when the constraints of the optimal control problem are so tight that the cost function plays no role in determining the solution. For instance, we have an abnormal case when the optimal solution \(t\mapsto \bigl(x\opt(t), u\opt(t)\bigr)\) is ``isolated'' in the sense that there is no other solution satisfying the end-point constraints in the vicinity --- as measured by the supremum norm --- of the optimal solution.

	\subsection{Reachability}
		We recast the problem \eqref{e:opt reach problem} as an optimal control problem with a discontinuous cost function as follows: Since \(\norm{u_2}_{\Lp 0} = \tfin - \tinit - \int_{\tinit}^{\tfin} \indic{\{0\}}(u_2(s))\,\dd s\), the optimal control problem \eqref{e:opt reach problem} is equivalent to
		\begin{equation}
		\label{e:opt reach control problem}
		\begin{aligned}
			& \minimize_{u_1, u_2}
								&& \quad - \int_{\tinit}^{\tfin} \indic{\{0\}}(u_2(s))\,\dd s\\
			& \sbjto			&& \quad 
				\begin{cases}
					\dot z(t) = A z(t) + B u_1(t) u_2(t) \quad \text{for a.e.\ }t\in[\tinit, \tfin],\\
					z(\tinit) = \zinit\in\R^d, \quad z(\tfin) = \zfin\in\R^d,\\
					u_1:[\tinit, \tfin]\lra \admact\subset\R^m\text{ compact},\\
					u_2:[\tinit, \tfin]\lra\{0, 1\},\\
					u_1, u_2\text{ Lebesgue measurable},
				\end{cases}
		\end{aligned}
		\end{equation}

		Measurability of the instantaneous cost function in \eqref{e:opt reach control problem} follows from the fact that it is an indicator function of a closed set in \(\R^m\). Theorem \ref{t:Clarke extended PMP} applied to the optimal control problem \eqref{e:opt reach control problem} yields the following:
		\begin{thm}
		\label{t:main reach result}
			Consider the optimal control problem \eqref{e:opt reach control problem}. Assume that \(\tfin - \tinit\) is larger than the minimum time necessary to execute the transfer \(z(\tinit) = \zinit\) to \(z(\tfin) = \zfin\). Associated to every solution \([\tinit, \tfin]\ni t\mapsto \bigl(z\opt(t), u_1\opt(t), u_2\opt(t)\bigr)\) to \eqref{e:opt reach control problem} there exist an absolutely continuous map \([\tinit, \tfin]\ni t\mapsto p(t)\in\R^d\) and a scalar \(\eta = 0\) or \(1\), such that for a.e.\ \(t\in[\tinit, \tfin]\):
			\[
			\left\{
			\begin{aligned}
				& \dot z\opt(t) = A z\opt(t) + B u_1\opt(t) u_2\opt(t),\quad z\opt(\tinit) = \zinit,\; z\opt(\tfin) = \zfin,\\
				& \dot p(t) = - A\transp p(t),\\
				& u_1\opt(t) \in \argmax_{v_1\in\admact} \inprod{B\transp p(t)}{v_1},\\
				& u_2\opt(t) = \begin{cases}
					\begin{cases}
						1	& \text{if }\sup_{v_1\in\admact}\inprod{B\transp p(t)}{v_1} \ge 1,\\
						0	& \text{otherwise,}
					\end{cases}
					& \text{if \(\eta = 1\)},\\
					1 & \text{if \(\eta = 0\)}.
				\end{cases}
			\end{aligned}
			\right.
			\]
		\end{thm}

		Theorem \ref{t:main reach result} features a \(2d\)-dimensional ordinary differential equation for \(t\mapsto\bigl(z\opt(t), p(t)\bigr)\), and as such is a well-posed problem in view of the fact that there are \(2d\) boundary conditions --- the initial and final conditions of \(z\opt\).

		\begin{rem}
			{\rm 
			Note that in the abnormal case, i.e., when \(\eta = 0\), we have \(u_2\opt(t) \equiv 1\) in Theorem \ref{t:main reach result}. This situation may occur, e.g., when the time difference \(\tfin - \tinit\) is the minimum time needed to execute the transfer of states from \(\zinit\) to \(\zfin\); in this situation we must have \(u_2\opt(t) \equiv 1\) for the entire duration of the aforementioned execution.%
			}
		\end{rem}

		\begin{rem}
		\label{r:connection to maximally sparse}
			{\rm 
			In the normal case, i.e., \(\eta = 1\), the control \(u_1\opt u_2\opt\) may be regarded as the sparsest possible control to execute the reachability manoeuvre in Theorem \ref{t:main reach result}.%
			}
		\end{rem}

		\textit{Proof of Theorem \ref{t:main reach result}:}
			We employ Theorem \ref{t:Clarke extended PMP} to derive our assertions. Notice that the instantaneous cost function \(\Lambda\) in this case is solely dependent on the control, so continuous differentiability of \(\Lambda\) with respect to the space variable is automatically satisfied. The Hamiltonian function for the optimal control problem \eqref{e:opt reach control problem} is
			\begin{multline*}
				\R^d\times(\admact\times\{0, 1\})\times\R^d\ni \bigl(\xi, (\mu_1, \mu_2), p\bigr) \mapsto\\
				H^\eta\bigl(\xi, (\mu_1, \mu_2), p\bigr) \Let \inprod{p}{A\xi + B \mu_1\mu_2} + \eta\indic{\{0\}}(\mu_2)\in\R;
			\end{multline*}
			The nontriviality condition in \eqref{e:Clarke:nontriviality} translates to
			\[
				\bigl(\eta, p(t)\bigr)\neq (0, 0)\quad \text{for all }t\in[\tinit, \tfin].
			\]
			Since \(E\) is the singleton \((\zinit, \zfin)\) in our case, the limiting normal cone \(N^L_E(\zinit, \zfin)\) of \(E\) at the point \(\bigl(z(\tinit), z(\tfin)\bigr)\) is \(\R^d\times\R^d\), and therefore the transversality condition \eqref{e:Clarke:transversality} in our setting is given by
			\[
				\bigl(p(\tinit), -p(\tfin)\bigr)\in\R^d\times\R^d.
			\]
			In other words, the end-points of the adjoint are unconstrained. The adjoint equation in \eqref{e:Clarke:adjoint} is given by
			\[
				-\dot p(t) = \partial_x H^\eta\bigl(\boldsymbol\cdot, u_1\opt(t), u_2\opt(t), p(t)\bigr) (x\opt(t)) = A\transp p(t),
			\]
			with the absolutely continuous solution:
			\[
				p(t) = \epower{-(t - \tinit) A\transp}p(\tinit) \quad\text{for all }t\in[\tinit, \tfin].
			\]
			Since the minimum time needed to execute the transfer \(z(\tinit) = \zinit\) to \(z(\tfin) = \zfin\) is smaller than \(\tfin - \tinit\), the Hamiltonian maximization condition \eqref{e:Clarke:Hamiltonian max} is given by
			\begin{multline*}
				H^\eta\bigl(z\opt(t), u_1\opt(t), u_2\opt(t), p(t)\bigr)\\
				= \sup_{\substack{v_1\in\admact\\ v_2\in\{0, 1\}}} \left\{ \inprod{p(t)}{A z(t) + B v_1 v_2} + \eta \indic{\{0\}}(v_2) \right\},
			\end{multline*}
			where the supremum is attained in view of Weierstrass's theorem since the function on the right-hand side above is upper semicontinuous in \((v_1,v_2)\) and \(\admact\times\{0, 1\}\) is compact. We see at once that the order of maximization is irrelevant, and that the optimal controls are given by
			\[
				\bigl(u_1\opt(t), u_2\opt(t)\bigr) \in \argmax_{(v_1, v_2)\in\admact\times\{0, 1\}} \left\{ \inprod{B\transp p(t)}{v_1}v_2 + \eta\indic{\{0\}}(v_2)\right\}.
			\]
			In other words, if \(\eta = 1\), then for a.e.\ \(t\in[\tinit, \tfin]\),
			\begin{align*}
				u_1\opt(t) & \in \argmax_{v_1\in\admact} \inprod{B\transp p(t)}{v_1},\\
				u_2\opt(t) & = \begin{cases}
					1	& \text{if }\sup_{v_1\in\admact} \inprod{B\transp p(t)}{v_1} \ge 1,\\
					0	& \text{otherwise};
				\end{cases}
			\end{align*}
			if \(\eta = 0\), then for a.e.\ \(t\in[\tinit, \tfin]\),
			\[
				u_1\opt(t) \in \argmax_{v_1\in\admact} \inprod{B\transp p(t)}{v_1},\quad 
				u_2\opt(t) = 1.
			\]
			The assertion follows at once from the steps above.\placeqed

		In the particular case of the dimension of \(u_1\) being \(1\) and \(\admact = [-1, 1]\), we have the following simple formulas for the optimal control if \(\eta = 1\):
		\begin{equation}
		\label{e:1d formulas}
			u_1\opt(t) = \sgn\bigl(B\transp p(t)\bigr),\quad 
			u_2\opt(t) = \begin{cases}
				1	& \text{if }\abs{B\transp p(t)} \ge 1,\\
				0	& \text{otherwise}.
			\end{cases}
		\end{equation}

	\subsection{Linear quadratic performance}
		The optimal control problem \eqref{e:opt lq problem} is equivalent to
		\begin{equation}
		\label{e:opt lq control problem}
		\begin{aligned}
			& \minimize_{u_1, u_2}
							&& \int_{\tinit}^{\tfin}\Bigl( \tfrac{1}{2} \inprod{z(t)}{Q z(t)} + \tfrac{1}{2}\inprod{u_1(t)}{R u_1(t)}\\
							& && \qquad - \gamma \indic{\{0\}}(u_2(t))\Bigr)\,\dd t + \tfrac{1}{2}\inprod{z(\tfin)}{Q_f z(\tfin)} \\
			& \sbjto\;	\quad	&&  
				\begin{cases}
					\dot z(t) = A z(t) + B u_1(t) u_2(t) \quad \text{for a.e.\ }t\in[\tinit, \tfin],\\
					z(\tinit) = \zinit\in\R^d,\\
					u_1:[\tinit, \tfin]\lra\R^m,\\
					u_2:[\tinit, \tfin]\lra\{0, 1\},\\
					u_1, u_2 \text{ Lebesgue measurable},
				\end{cases}
		\end{aligned}
		\end{equation}
		where \(\gamma > 0\) is a fixed constant.

		Measurability of the instantaneous cost function follows from the fact that the indicator function is one of a closed set in \(\R^m\). Theorem \ref{t:Clarke extended PMP} applied to the optimal control problem \eqref{e:opt lq control problem} yields the following:
		\begin{thm}
		\label{t:main lq result}
			In the optimal control problem \eqref{e:opt lq control problem}, associated to every solution \([\tinit, \tfin]\ni t\mapsto \bigl(z\opt(t), u_1\opt(t), u_2\opt(t)\bigr)\) to \eqref{e:opt lq control problem} there exists an absolutely continuous map \([\tinit, \tfin]\ni t\mapsto p(t)\in\R^d\), such that for a.e.\ \(t\in[\tinit, \tfin]\):
			\[
			\left\{
			\begin{aligned}
				& \dot z\opt(t) = A z\opt(t) + B R\inverse B\transp p(t) u_2\opt(t),\quad z\opt(\tinit) = \zinit,\\
				& \dot p(t) = Q z\opt(t) - A\transp p(t),\quad p(\tfin) = - Q_f z\opt(\tfin),\\
				& u_1\opt(t) = \begin{cases}
					R\inverse B\transp p(t)	& \text{if }u_2\opt(t) = 1,\\
					0	& \text{otherwise},
				\end{cases}\\
				& u_2\opt(t) = \begin{cases}
					1	& \text{if }\inprod{B\transp p(t)}{R\inverse B\transp p(t)} \ge \gamma,\\
					0	& \text{otherwise}.
				\end{cases}
			\end{aligned}
			\right.
			\]
		\end{thm}

		Theorem \ref{t:main lq result} features a \(2d\)-dimensional ordinary differential equation for \(t\mapsto \bigl(z\opt(t), p(t)\bigr)\) with \(2d\) boundary conditions --- initial condition for \(z\opt\) and final condition for \(p\). As such it is a well-posed problem.

		\begin{rem}
			{\rm 
			Unlike in the reachability manoeuvres treated above, the abnormal case (\(\eta = 0\)) does not arise in the setting of Theorem \ref{t:main lq result}, as we shall establish in the proof of Theorem \ref{t:main lq result} given below.%
			}
		\end{rem}

		\begin{rem}
			{\rm 
			Note that the optimal control \(u_1\opt\) is sparse in the sense that it is set to `off' or \(0\) at certain times. It is in fact the sparsest control that strikes a balance between the \(\Lp 2([\tinit, \tfin])\) costs corresponding to the states and control \(u_1\) versus the \(\Lp 0([\tinit, \tfin])\) costs corresponding to the signal \(u_2\).%
			}
		\end{rem}

		\textit{Proof of Theorem \ref{t:main lq result}:}
			We employ Theorem \ref{t:Clarke extended PMP} to derive our assertions. Notice that the instantaneous cost function \(\Lambda\) in this case depends quadratically on the states and on the control \(u_1\) in addition to the \(\Lp 0([\tinit, \tfin])\) seminorm of the signal \(u_2\), so continuous differentiability of \(\Lambda\) with respect to the space variable is satisfied. The Hamiltonian function corresponding to the optimal control problem \eqref{e:opt lq control problem} is given by
			\begin{multline*}
				\R^d\times(\R^m\times\{0, 1\})\times\R^d\ni \bigl(\xi, (\mu_1, \mu_2), p\bigr) \mapsto \\
				H^\eta\bigl(\xi, (\mu_1, \mu_2), p\bigr)\Let \inprod{p}{A\xi + B \mu_1 \mu_2} \\
				- \eta\bigl(\tfrac{1}{2}\inprod{\xi}{Q\xi} + \tfrac{1}{2} \inprod{\mu_1}{R \mu_1} - \gamma\indic{\{0\}}(\mu_2)\bigr)\in\R.
			\end{multline*}
			The nontriviality condition \eqref{e:Clarke:nontriviality} translates to the condition
			\[
				\bigl(\eta, p(t)\bigr)\neq (0, 0)\quad\text{for all }t\in[\tinit, \tfin].
			\]
			Since the final cost function \(\ell\bigl(z(\tfin)\bigr) = \tfrac{1}{2}\inprod{z(\tfin)}{Q_f z(\tfin)}\) is smooth, the object \(\partial_x\ell\) is precisely the gradient \(Q_f z(\tfin)\) of \(\ell\), and since the constraint at the final time is absent, the transversality condition \eqref{e:Clarke:transversality} becomes
			\[
				-p(\tfin) = \eta Q_f z\opt(\tfin).
			\]
			The Hamiltonian is smooth in the space variable \(\xi\); consequently, the adjoint equation \eqref{e:Clarke:adjoint} is given by
			\begin{align*}
				- \dot p(t) & = \partial_\xi H^\eta\bigl(\pmb{\cdot}, u_1\opt(t), u_2\opt(t), p(t)\bigr)\bigl(z\opt(t)\bigr)\\
					& = A\transp p(t) - \eta Q z\opt(t)\quad \text{a.e.\ }t\in[\tinit, \tfin].
			\end{align*}
			To wit, the adjoint equation is the following boundary value problem:
			\[
			\left\{
			\begin{aligned}
				& \dot p(t) = - A\transp p(t) + \eta Q z\opt(t)\quad \text{a.e.\ }t\in[\tinit, \tfin],\\
				& p(\tfin) = - \eta Q_f z\opt(\tfin).
			\end{aligned}
			\right.
			\]
			We claim that \(\eta = 1\). Indeed, if not, then the terminal boundary condition \(-\eta Q_f z\opt(\tfin)\) and the forcing term \(\eta Q z\opt(t)\) in the adjoint equation both vanish. In view of the resulting linearity of the adjoint equation, the entire map \([\tinit, \tfin]\ni t\mapsto p(t)\in\R^d\) vanishes. But then this contradicts the nontriviality condition mentioned above. The adjoint equation is, therefore, given by
			\[
			\left\{
			\begin{aligned}
				& \dot p(t) = - A\transp p(t) + Q z\opt(t)\quad \text{a.e.\ }t\in[\tinit, \tfin],\\
				& p(\tfin) = - Q_f z\opt(\tfin).
			\end{aligned}
			\right.
			\]

			In view of the preceding analysis we commit to \(\eta = 1\) and henceforth write \(H\) instead of \(H^\eta\). We have
			\begin{multline*}
				H\bigl(\xi, (\mu_1, \mu_2), p\bigr) = \inprod{p}{A\xi + B \mu_1 \mu_2} + \gamma \indic{\{0\}}(\mu_2)\\
					- \tfrac{1}{2}\bigl(\inprod{\xi}{Q\xi} + \inprod{\mu_1}{R \mu_1}\bigr).
			\end{multline*}
			Observe that 
			\begin{itemize}[label=\(\circ\), leftmargin=*]
				\item the order of maximization of the Hamiltonian function \(H\) with respect to the controls is irrelevant;
				\item the Hamiltonian function \(H\) is smooth and concave in \(u_1\) on \(\R^m\) due to positive definiteness of the matrix \(R\), which shows that the maximum over \(u_1\in\R^m\) is unique;
				\item the Hamiltonian function \(H\) is upper semicontinuous in \(u_2\), and by Weierstrass's theorem the maximum is attained on the compact set \(\{0, 1\}\).
			\end{itemize}
			Therefore, the Hamiltonian maximization condition \eqref{e:Clarke:Hamiltonian max} leads to: for a.e.\ \(t\in[\tinit,\tfin]\),
			\begin{align*}
				& \bigl(u_1\opt(t), u_2\opt(t)\bigr) \in \\
				& \qquad \argmax_{(v_1, v_2)\in\R^m\times\{0, 1\}} \left\{ \inprod{B\transp p(t)}{v_1} v_2 - \tfrac{1}{2}\inprod{v_1}{Rv_1} + \gamma \indic{\{0\}}(v_2)\right\},
			\end{align*}
			which gives
			\[
				u_1\opt(t) = \begin{cases}
					R\inverse B\transp p(t)	& \text{if }u_2\opt(t) = 1,\\
					0						& \text{otherwise},
				\end{cases}
			\]
			and
			\[
				u_2\opt(t) = \begin{cases}
					1	& \text{if }\inprod{B\transp p(t)}{u_1\opt(t)} \ge \gamma,\\
					0	& \text{otherwise}.
				\end{cases}
			\]
			The assertion follows at once from the steps above.\placeqed

		From Theorem \ref{t:main lq result} we get the following `canonical' set of dynamical equations, in which the matrix \(\mcal H(t)\) is sometimes referred to as the Hamiltonian matrix:
		\begin{align*}
			\pmat{\dot z\opt(t)\\ \dot p(t)} & = \pmat{A & BR\inverse B\transp u_2\opt(t)\\ Q & -A\transp} \pmat{z\opt(t)\\ p(t)}\\
				& \teL \mcal H(t) \pmat{z\opt(t)\\ p(t)}\quad\text{for a.e.\ }t\in[\tinit, \tfin].
		\end{align*}
		Letting \(S \Let \left\{s\in[\tinit, \tfin]\,\Big|\, \inprod{B\transp p(s)}{R\inverse B\transp p(s)} \ge \gamma\right\}\), we rewrite the Hamiltonian matrix as
		\[
			\mcal H(t) = \pmat{A & BR\inverse B\transp \indic{S}(t)\\ Q & -A\transp}.
		\]
		Standard arguments as in \cite[Chapter 6]{ref:Lib-12} may be employed to show that the state adjoint \(p(t)\) is linearly related to \(z\opt(t)\) given by \(p(t) = - P(t) z\opt(t)\), where \(P(\cdot)\) satisfies the ordinary differential equation
		\begin{multline}
		\label{e:pulsating Riccati}
			\dot P(t) + A\transp P(t) + P(t) A + Q \\
			- P(t) B R\inverse B\transp P(t) \indic{S}(t) = 0\quad \text{for a.e. }t\in[\tinit, \tfin],
		\end{multline}
		with boundary condition \(P(\tfin) = Q_f\). This \emph{Riccati} equation \eqref{e:pulsating Riccati} is a bona fide ``hybrid'' ordinary differential equation; to our knowledge no closed form solution to this differential equation is available. It switches between a Lyapunov equation and a full-fledged Riccati differential equation at time \(s\in[\tinit, \tfin]\) depending on whether \(\norm{p(s)}_{BR\inverse B\transp}^2 \ge \gamma\) or not, where \(\norm{\cdot}_M \Let \sqrt{\inprod{\cdot}{M\cdot}}\) for a symmetric and non-negative definite matrix \(M\). Note that \eqref{e:pulsating Riccati} is intimately connected with the dynamics of the states \(z\opt(\cdot)\), which makes it a challenging equation to deal with.

		\begin{rem}
		\label{r:sparse quadratic regulator}
			{\rm 
			Consider the quadratic regulator problem with \(\Lp 0\)-regularization:
			\begin{equation}
			\label{e:sparse lq}
			\begin{aligned}
				& \minimize_{u}
								&& \int_{\tinit}^{\tfin}\Bigl( \tfrac{1}{2} \inprod{z(t)}{Q z(t)} + \tfrac{1}{2}\inprod{u(t)}{R u(t)}\\
								& && \qquad - \gamma \indic{\{0\}}(u(t))\Bigr)\,\dd t + \tfrac{1}{2}\inprod{z(\tfin)}{Q_f z(\tfin)} \\
				& \sbjto\;	\quad	&&  
					\begin{cases}
						\dot z(t) = A z(t) + B u(t) \quad \text{for a.e.\ }t\in[\tinit, \tfin],\\
						z(\tinit) = \zinit\in\R^d,\\
						u:[\tinit, \tfin]\lra\R^m,\\
						u\text{ Lebesgue measurable},
					\end{cases}
			\end{aligned}
			\end{equation}
			where \(\gamma > 0\) is a fixed constant. If \((z\opt, u\opt)\) is an optimal state-action trajectory solving \eqref{e:sparse lq}, observe that \(u\opt\) is by definition sparsest in the sense that it is turned off for the maximal duration of time; cf.\ \cite{ref:JovLin-13}. Straightforward calculations with the support of the nonsmooth Pontryagin maximum principle Theorem \ref{t:Clarke extended PMP} shows that the optimal control for the problem \eqref{e:sparse lq} is characterized by
			\[
				u\opt(t) = \begin{cases}
					R\inverse B\transp p(t)	& \text{if }\norm{p(t)}_{BR\inverse B\transp}^2 \ge \gamma,\\
					0	& \text{otherwise},
				\end{cases}
			\]
			where \([\tinit, \tfin]\ni t\mapsto p(t)\in\R^d\) is an absolutely continuous map that solves the differential equation
			\begin{align*}
				\begin{cases}
					\dot p(t) = - A\transp p(t) + Q z\opt(t),\\
					p(\tfin) = - Q_f z\opt(\tfin).
				\end{cases}
			\end{align*}
			The abnormal case (\(\eta = 0\)) does not arise here, as can be readily seen by mimicking the arguments in the proof of Theorem \ref{t:main lq result}.%
			}
		\end{rem}

		\begin{rem}
			{\rm 
			It remains a challenging open problem to ensure stability of the closed-loop system in the \(\Lp 0\)-regularized LQ problem discussed in Remark \ref{r:sparse quadratic regulator}. The standard analysis of letting the final time \(\tfin \to+\infty\) and analyzing the associated Riccati equation turns out to be difficult because the Riccati equation in this setting becomes hybrid, with a discontinuity set connected to the dynamics of the adjoint \(p\).
			}
		\end{rem}

	\subsection{Existence of optimality}
		So far we have employed necessary conditions for solutions to \eqref{e:opt reach problem} and \eqref{e:opt lq problem} under the aegis of a nonsmooth Pontryagin maximum principle, but have sidestepped the matter of sufficient conditions for optimality of the state-action trajectories satisfying the necessary conditions. In this subsection we treat the problem of optimality of such state-action trajectories. In other words, having identified the extremals corresponding to the problems \eqref{e:opt reach problem} and \eqref{e:opt lq problem}, we wish to ascertain whether the necessary conditions in Theorem \ref{t:main reach result} and Theorem \ref{t:main lq result} are also sufficient for optimality.

		To this end, we have the following:
		\begin{prop}
		\label{p:existence}
			\mbox{}
			\begin{enumerate}[label={\rm (\ref{p:existence}-\alph*)}, leftmargin=*, widest=b, align=left]
				\item \label{p:existence:reach} Consider the problem \eqref{e:opt reach problem}. If there exist adjoint solutions to \eqref{e:opt reach problem} corresponding to \(\eta = 1\) satisfying the conditions of Theorem \ref{t:main reach result}, then the corresponding state-action trajectories \([\tinit, \tfin]\ni t\mapsto \bigl(z\opt(t), u_1\opt(t), u_2\opt(t)\bigr)\) are locally optimal.
				\item \label{p:existence:lq} For the problem \eqref{e:opt lq problem} state-action trajectories satisfying the conditions in Theorem \ref{t:main lq result} are locally optimal.
			\end{enumerate}
		\end{prop}
		\textit{Proof:}
			\ref{p:existence:reach}: Assume that \(\eta = 1\), and let \(t\mapsto \bigl(z\opt(t), u_1\opt(t), u_2\opt(t)\bigr)\) denote a state-action trajectory satisfying the corresponding assertions of Theorem \ref{t:main reach result}. Pick \(\delta > 0\), and note that the map
			\[
				\R^d\ni z\mapsto \Bigl( \inprod{p(t)}{A z + B u_1\opt(t) u_2\opt(t)} + \indic{\{0\}}(u_2\opt(t)) \Bigr)\in\R
			\]
			is affine and therefore concave on the set 
			\[
				\Bigl\{z\in\R^d\,\Big|\, \norm{z - z\opt(t)} < \delta\Bigr\} \quad\text{for a.e.\ }t\in[\tinit, \tfin].
			\]
			Now \cite[Corollary 24.2]{ref:Cla-13} applies directly, and implies that \(t\mapsto \bigl(z\opt(t), u_1\opt(t), u_2\opt(t)\bigr)\) is optimal in the \(\delta\)-neighborhood of \(z\opt(\cdot)\).

			\ref{p:existence:lq}: Let \(t\mapsto\bigl(z\opt(t), u_1\opt(t), u_2\opt(t)\bigr)\) denote a state-action trajectory satisfying the assertions of Theorem \ref{t:main lq result}. Pick \(\delta > 0\). Assume that the conditions in Theorem \ref{t:main lq result} hold. Then the map
			\begin{align*}
				\R^d\ni z\mapsto \biggl( & \inprod{p(t)}{Az + B u_1\opt(t) u_2\opt(t)} + \gamma \indic{\{0\}}(u_2\opt(t)) \\
					& - \tfrac{1}{2}\Bigl(\inprod{z}{Qz} + \inprod{u_1\opt(t)}{R u_1\opt(t)}\Bigr) \biggr)\in\R
			\end{align*}
			is concave on the set \(\Bigl\{z\in\R^d\,\Big|\, \norm{z - z\opt(t)} < \delta\Bigr\}\) for a.e.\ \(t\in[\tinit, \tfin]\). Once again, \cite[Corollary 24.2]{ref:Cla-13} immediately gives us optimality of \(t\mapsto \bigl(z\opt(t), u_1\opt(t), u_2\opt(t)\bigr)\) in the \(\delta\)-neighborhood of \(z\opt(\cdot)\).\placeqed

	\section{Examples}
	\label{s:examples}


\begin{exmp}
{\rm 
Now we illustrate the optimal control problem on the Linear Quadratic performance index in the presence of the jammer, problem \eqref{e:opt lq problem}. The first set of simulations consider a linearized, second-order inverted pendulum dynamics as below,
\begin{align}
A &= \begin{pmatrix}
0 & 1 \\
\frac{mgl}{I} & \frac{-b}{I}
\end{pmatrix} \quad B = \begin{pmatrix}
0 \\
1
\end{pmatrix}
\end{align}
To the aforementioned dynamical system, the optimal control described by Theorem~\ref{t:main lq result} is applied with parameter values, $ m = 2\, kg,\,l =1\, m,g = 9.81\, m/s^2,\,I = ml^2/3\, kgm^2,\,b = 0.02$, weights, $Q = \begin{pmatrix}
3 & 0 \\
0 & 3
\end{pmatrix},\,Q_f = \begin{pmatrix}
10 & 0\\
0 & 10
\end{pmatrix},\, R = 3,\, \gamma = 0.01$, and initial conditions $x_0 = \pmat{0\\ \pi/10}$. The integration tolerance for all cases is kept at 1e-4.

\begin{figure}
\centering
	\includegraphics[width=0.45\textwidth]{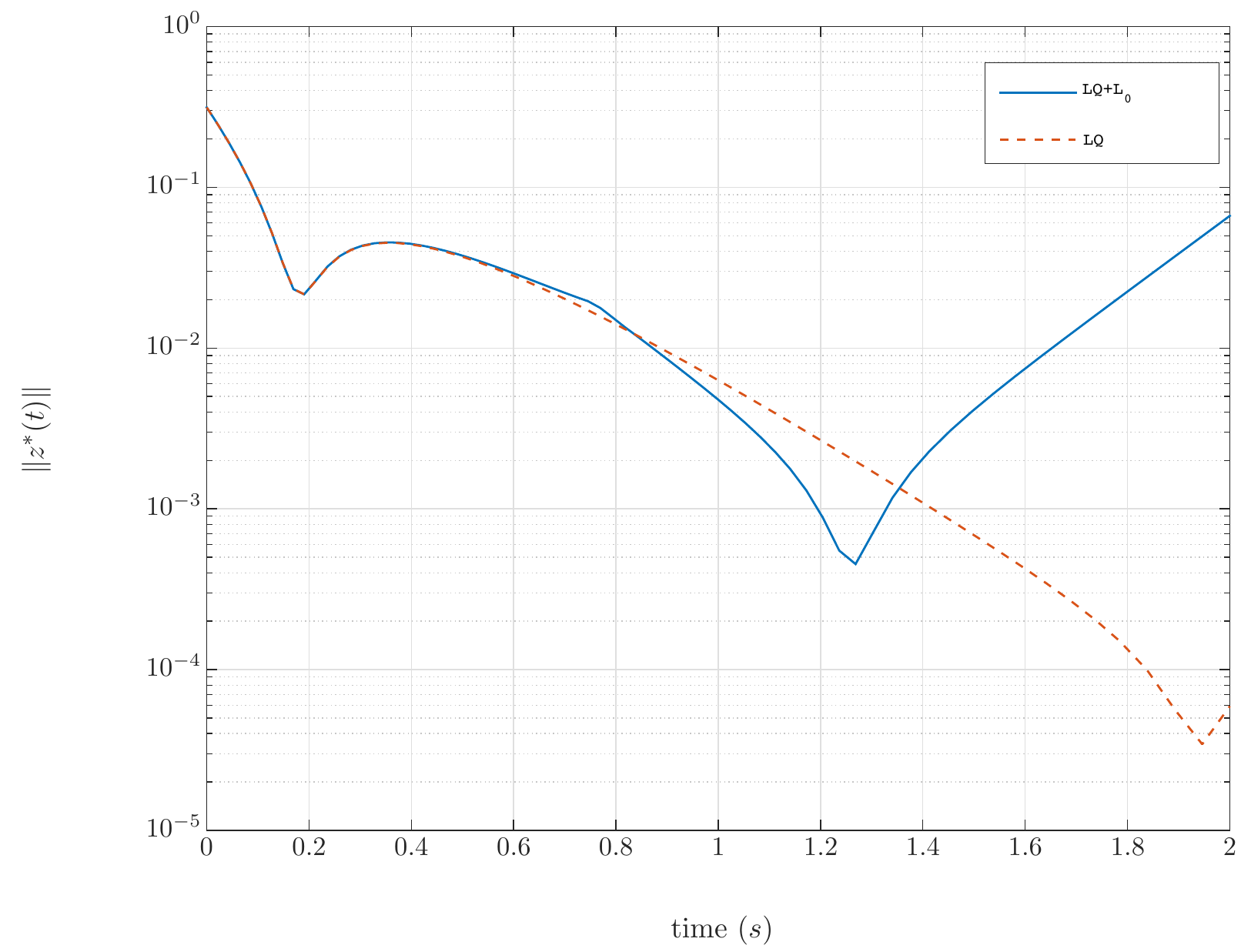}
	\captionof{subfigure}{Norm of states \(\norm{z\opt(t)}\) against \(t\)}
	\includegraphics[width=0.45\textwidth]{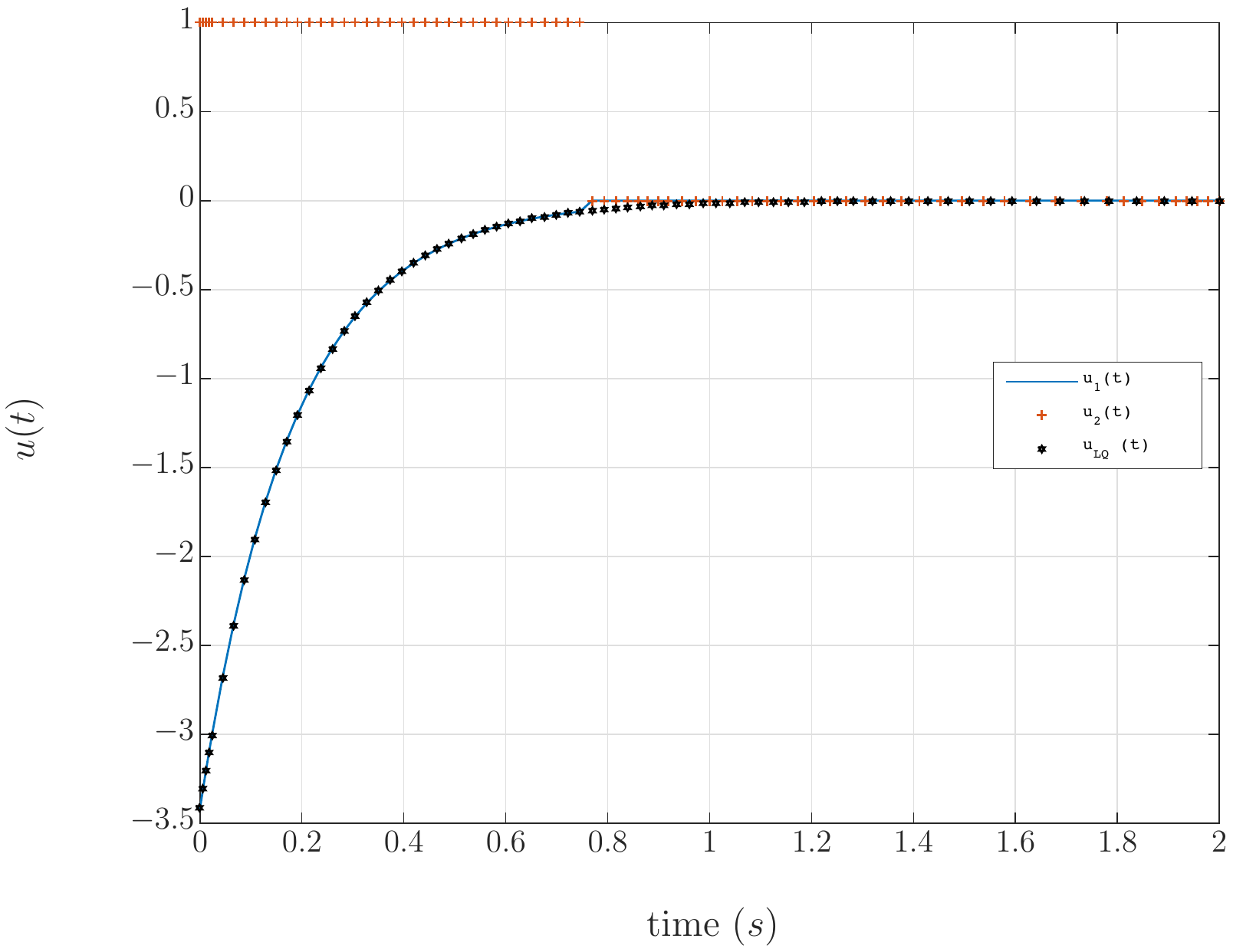}
	\captionof{subfigure}{Control \(u_1\opt(t)\) and jammer \(u_2\opt(t)\) against \(t\)}
	\captionof{figure}{Inverted pendulum stabilization with $\gamma = 0.01$}
\label{fig:LQ}
\end{figure}

%
The two point boundary value problem (TPBVP) arising from Theorem \ref{t:main lq result} is solved using a multiple-shooting technique \cite{ref:Bet-09}. The aim of multiple shooting is to iterate on an appropriate value of $p(\bar{t})$ such that given an initial condition, $z^*(\bar{t})$, the final constraint, $p(\tfin) = - Q_f z\opt(\tfin)$ is satisfied. The iterates are computed using a suitable nonlinear programming (NLP) technique. The current article utilizes the trust-region based \textsf{fmincon} routine in MATLAB$^\copyright$. A comparison of numerical efficiency of NLP schemes can be found in \cite{ref:Benson-03}, \cite{ref:SchitZillZote-94}, \cite{ref:BetEldHuf-93}. The simulated results for a time span of $2\,s$ are shown in Figure~\ref{fig:LQ}. The plots show the evolution of $\|z\opt(t)\|$ as well as the commanded control $u_1\opt(t)$ and jammer $u_2\opt(t)$. The jammer signal $u_2\opt(t)$ goes to zero approximately beyond $0.75\,s$ as evident from the plots. For the given set of parameter values, initial conditions and weights this indicates the maximum duration over which the control can be switched off while still optimizing the prescribed modified Linear Quadratic performance \eqref{e:opt lq control problem}. The plot of $\|z\opt(t)\|$ shows a clear decay to 4e-4 before starting to rise again. $\|z\opt(t)\|$ continues to decay well beyond $0.77 \, s$ when $u_2\opt(t)$ goes to zero and starts to rise again under the influence of unstable dynamics beyond $t = 1.25 \, s$. In Figure~\ref{fig:LQ}, is also superimposed the optimal trajectories corresponding to the classical Linear Quadratic Regulator (LQR). The corresponding trajectories are obtained simply by setting $\gamma = 0$ in the optimal control problem~(\ref{e:opt lq control problem}). As expected, $\|z\opt(t)\|$ corresponding to the classical LQR solution converges to about 2e-5, which is much lower than our non-smooth solution based on Theorem~\ref{t:main lq result}. It is however interesting to note that at around $t = 0.77 \, s$,  $\|z\opt(t)\|$ corresponding to implementation of Theorem~\ref{t:main lq result} ($\gamma = 0.01$) starts to decay at a faster rate than the classical LQR case ($\gamma = 0$). The sudden increase in the decay rate is coincident with deviation of $u_1\opt(t)$ from $u_{LQ}$ corresponding to the classical LQR solution. The deviation in the control magnitudes for both cases lasts for about $0.02 \, s$ beyond which $u_1\opt(t) = 0$ while $u_{LQ}$ continues to asymptotically converge to zero.
\begin{figure}
\centering
	\includegraphics[width=0.45\textwidth]{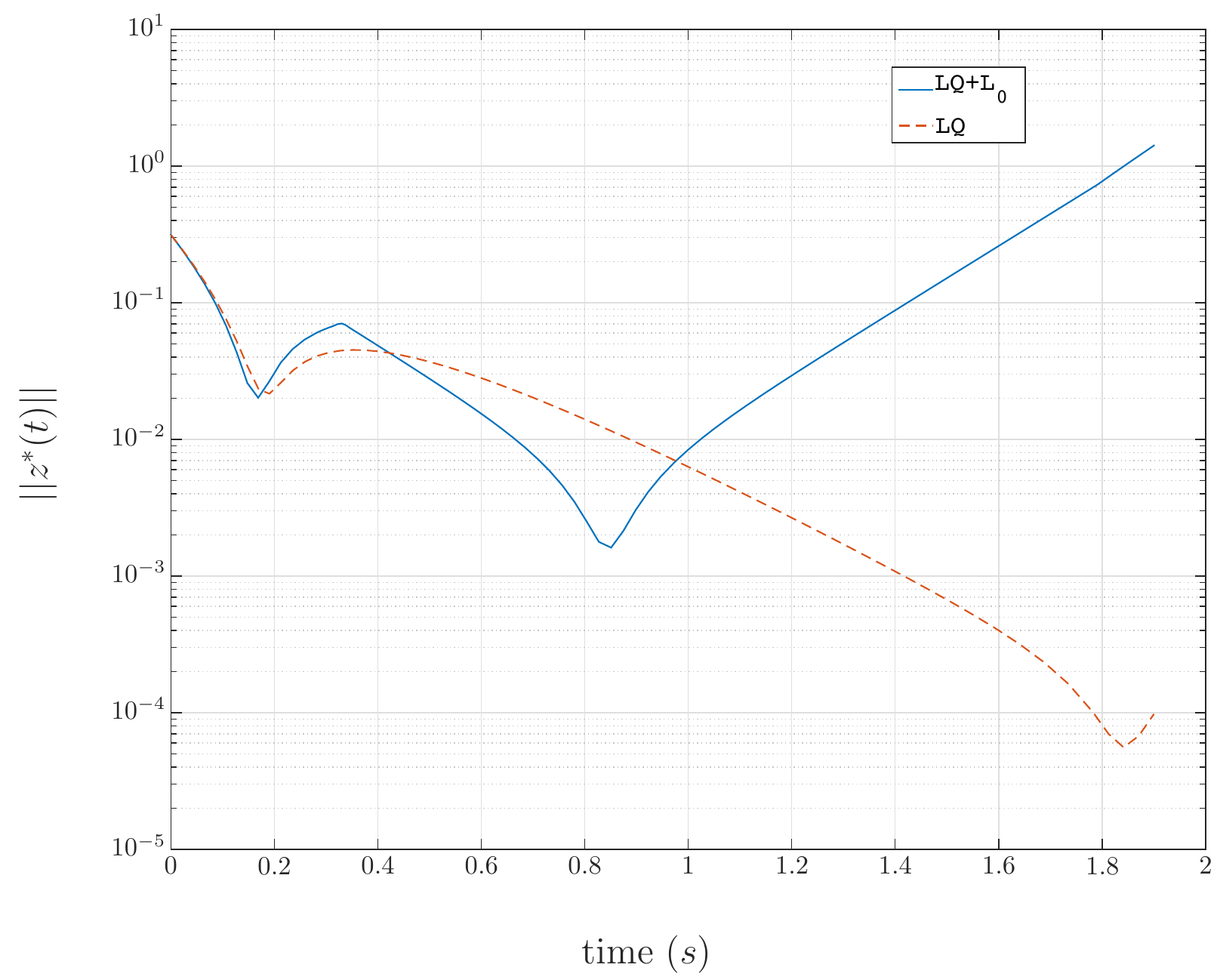}
	\captionof{subfigure}{Norm of states \(\norm{z\opt(t)}\) against \(t\)}
	\includegraphics[width=0.45\textwidth]{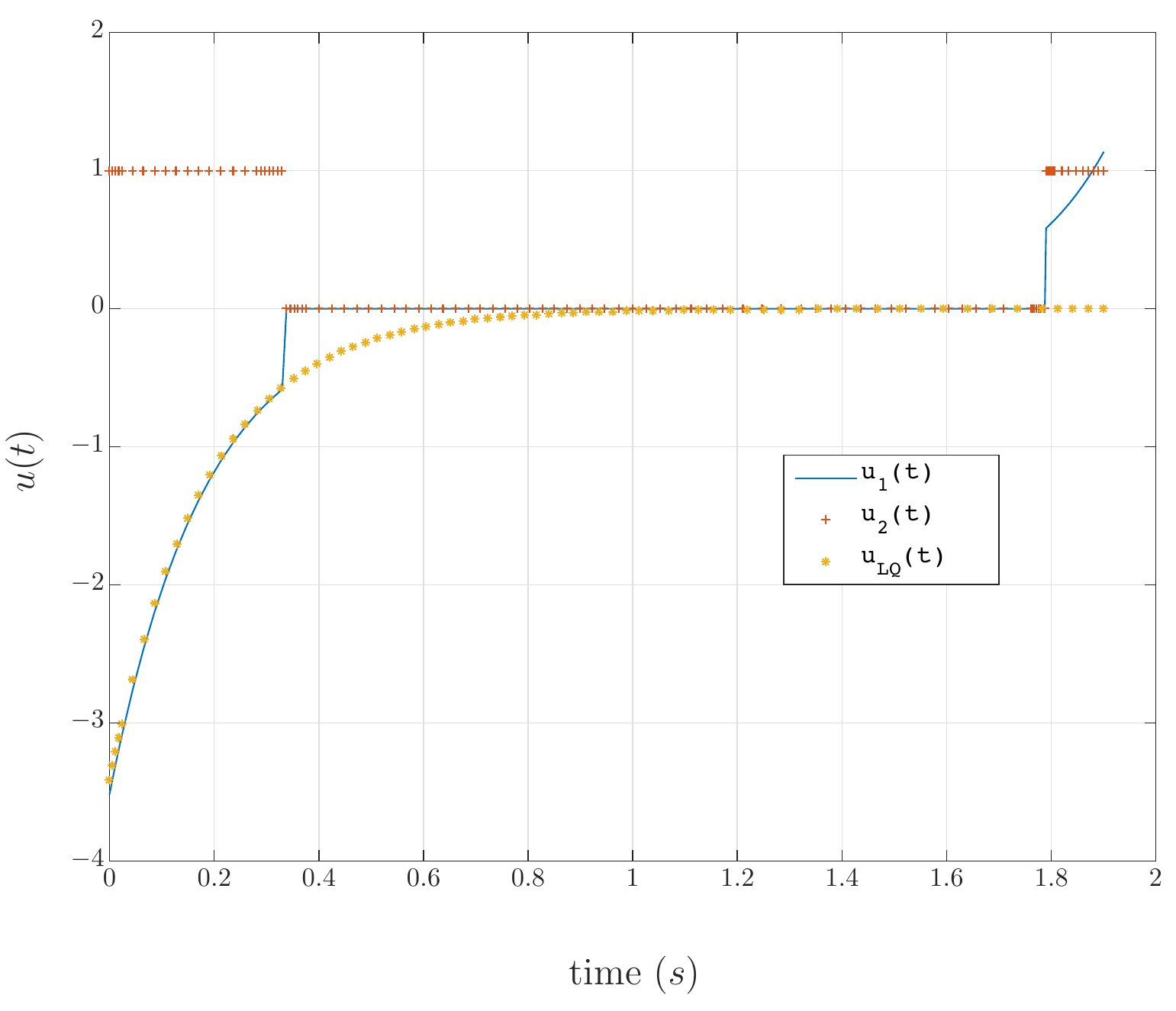}
	\captionof{subfigure}{Control \(u_1\opt(t)\) and jammer \(u_2\opt(t)\) against \(t\)}
	\captionof{figure}{Inverted pendulum stabilization example with $\gamma=1$}
\label{fig:LQ_rolling}
\end{figure}

In order to illustrate the effect of L$_0$ cost on the jammer, another set of simulations with $\gamma = 1$ is shown in Figure~\ref{fig:LQ_rolling} along with the classical LQR control solution. Similar to the $\gamma = 0.01$ case shown in Figure~\ref{fig:LQ}, a distinct change in the control magnitude is observed at around $0.35 s$ which also corresponds to faster rate of decay of $\|z^*(t)\|$ with the LQ$+$L$_0$ based control from $0.35-0.85 \, s$. However, as expected, a higher weightage on the L$_0$ norm of the jammer results in longer span of time with $u_2^*(t) = 0$ ($\approx$1.45 $s$), as compared to the previous case with $\gamma = 0.01$ ($\approx$1.15 $s$). On the contrary, the least value achieved by $\|z^*(t)\|$ is 2e-3 when $\gamma = 1$, while it is 4e-4 for the $\gamma = 0.01$ case. These differences are due to changes in the relative weightage of each term in the cost (\ref{e:opt lq control problem}).
}
\end{exmp}

\begin{exmp}
{\rm 
For the next set of simulations, a linearized inverted pendulum on a cart system is considered. The fourth order model is represented by,
\begin{align}
A = \begin{pmatrix}
0 & 1 & 0 & 0\\
0 & 0 & -\frac{mg}{M} & 0 \\
0 & 0 & 0 & 1 \\
0 & 0 & \frac{(m+M)g}{(2Ml)} & 0
\end{pmatrix} \quad B = \begin{pmatrix}
0 \\ 
\frac{1}{M} \\
0 \\
-\frac{1}{(2Ml)}
\end{pmatrix}.
\end{align}
The optimal control as per Theorem~\ref{t:main lq result} is computed as in the second order example for parameter values, $ m = 2\,kg,\,l =1\, m,g = 9.81 \,m/s^2, \,M = 2 \,kg$, weights, $Q = \begin{pmatrix}
1 & 0 & 0 & 0 \\
0 & 1 & 0 & 0 \\
0 & 0 & 1 & 0 \\
0 & 0 & 0 & 1
\end{pmatrix},\,Q_f = 100, \,\, R = 1,\, \gamma = 0.1$ and initial conditions, $x_0 = \pmat{0 & \pi/10 & 0 & 0}\transp$. Figure~\ref{fig:LQ4} shows the plot of $\|z\opt(t)\|$ evolution with the optimal control, $u_1\opt(t)$ and jammer, $u_2\opt(t)$ for a time span of $1.9 \, s$. 

\begin{figure}
\centering
	\includegraphics[width=0.45\textwidth]{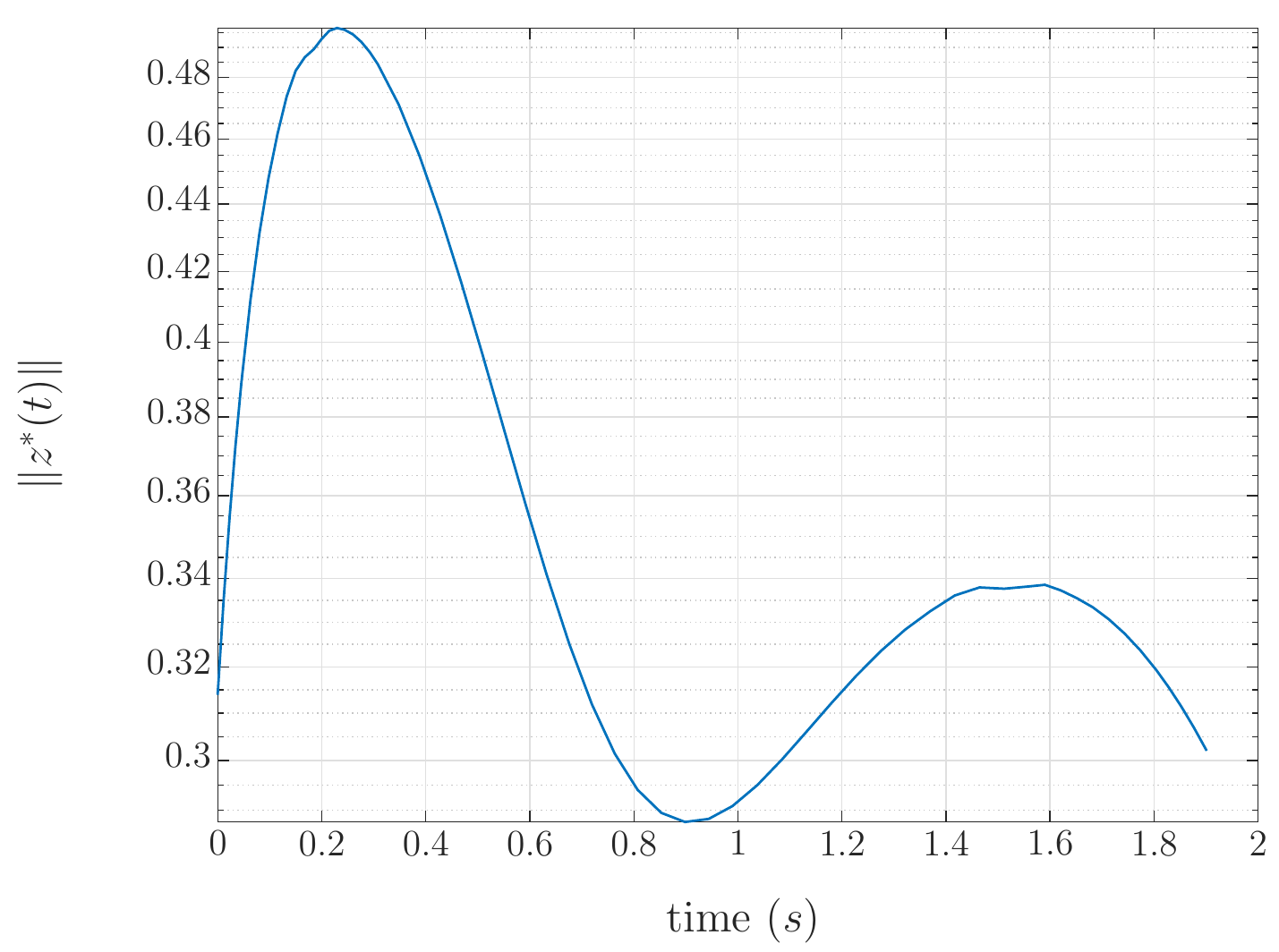}
	\captionof{subfigure}{Norm of states \(\norm{z\opt(t)}\) against \(t\)}
	\includegraphics[width=0.45\textwidth]{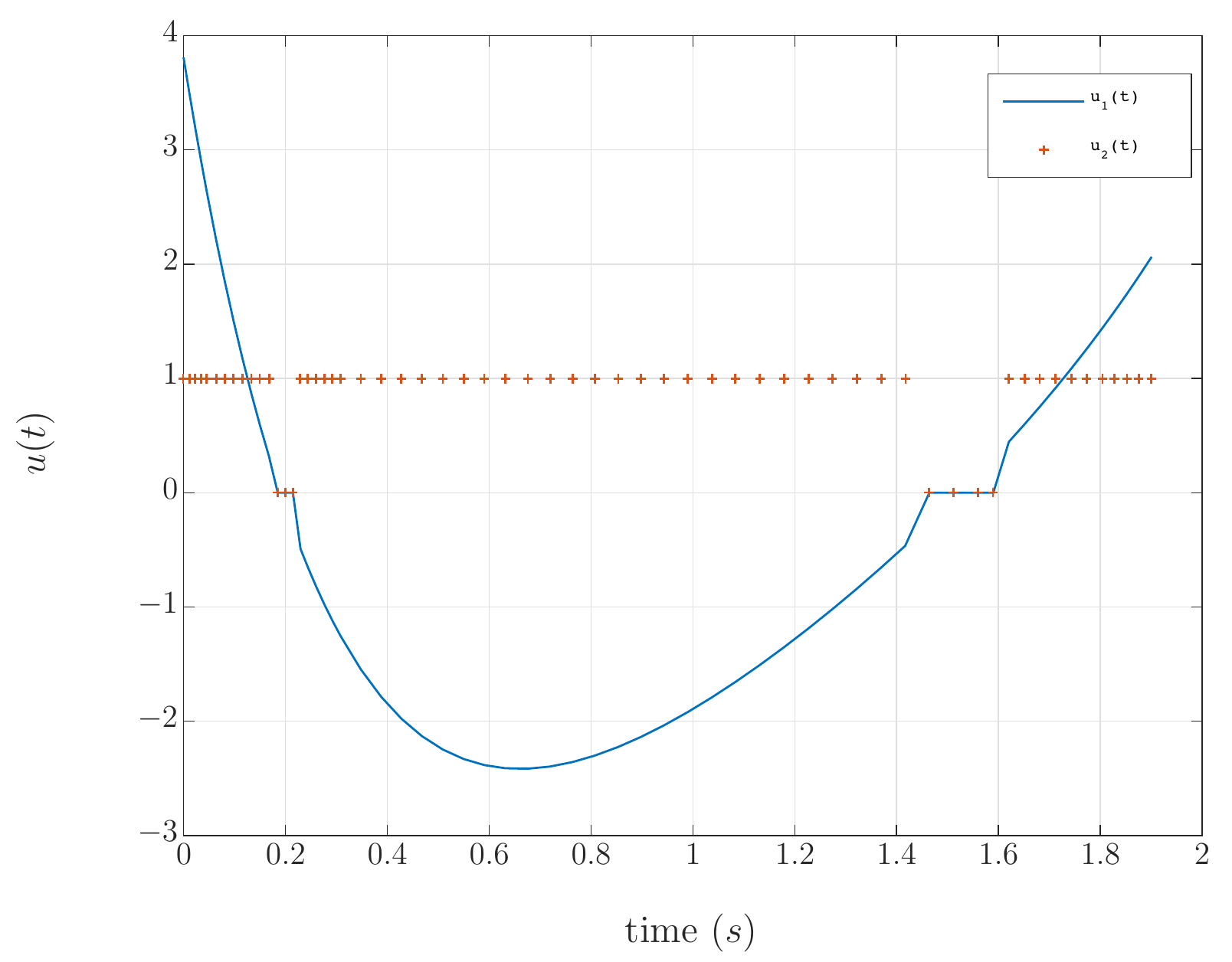}
	\captionof{subfigure}{Control \(u_1\opt(t)\) and jammer \(u_2\opt(t)\) against \(t\)}
	\captionof{figure}{Inverted pendulum on cart stabilization}
\label{fig:LQ4}
\end{figure}



The optimal jammer signal, $u_2\opt(t)$ is initially non-zero and goes intermittently to zero for a short time span around $0.2 \, s$ and $1.5 \, s$ indicating zero control input to the system. On careful examination of the $\|z\opt(t)\|$ plot, the phase of zero control is reflected in the form of sharp changes in the norm. After a period of initial decay up to around $t = 0.9 \,s$, $\|z\opt(t)\|$ rises again. Compared to the second order case, the controls $u_1\opt(t)u_2\opt(t)$ are required to be `on' for a larger percentage of the simulation window as observed from the plots.
}
\end{exmp}


	\section{Conclusion}
	\label{s:conclusion}
		We have studied the reachability problem \eqref{e:opt reach problem} and the LQ optimal control problem \eqref{e:opt lq problem}, both in the presence of a jammer, and have derived necessary and sufficient conditions for optimality in \secref{s:main results}; our primary analytical apparatus was a non-smooth Pontryagin maximum principle. In \secref{s:examples} we have compared the performance of the linear quadratic problem in the presence of a jammer against its standard operation.

	\begin{ack}
		The authors thank Harish Pillai and Debasattam Pal for helpful discussions on the Riccati equation. S.\ Srikant was supported in part by the grant 12IRCCSG007 from IRCC, IIT Bombay. D.\ Chatterjee was supported in part by the grant 12IRCCSG005 from IRCC, IIT Bombay.
	\end{ack}

\end{document}